\documentclass{aa}
\usepackage{graphicx}
\usepackage{txfonts}
\begin{document}
   \title{Gaussian decomposition of \ion{H}{i} surveys}
   \subtitle{IV. Galactic intermediate- and high-velocity clouds}
   \author{U. Haud}
   \institute{Tartu Observatory, 61\,602 T\~oravere, Tartumaa, Estonia\\
              \email{urmas@aai.ee} }
   \date{Received \today; accepted \today}
   \abstract
      {Traditionally intermediate- (IVC) and high-velocity hydrogen
       clouds (HVC) were defined to be concentrations of \ion{H}{i} gas,
       with line-of-sight velocities that are inconsistent with data on
       the differential rotation of the Galaxy.}
      {We attempt to demonstrate that IVCs and HVCs can be identified
       from density enhancements in parameter distributions of Galactic
       \ion{H}{i} 21 cm radio lines.}
      {To investigate the properties of the 21 cm radio lines, the
       profiles of ``The Leiden/Argentine/Bonn (LAB) Survey of Galactic
       \ion{H}{i}'' are decomposed into Gaussian components using a
       fully automatic algorithm. We focus on some regions with an
       increased number of Gaussians in phase space, defined by the
       component central velocity ($V_\mathrm{C}$) and the full width at
       the level of half maximum (FWHM). To separate the Gaussians
       responsible for the phase-space density enhancements, we model
       the width distributions of Gaussian components at
       equally-populated velocity intervals, using lognormal
       distributions.}
      {We study the Gaussians, which parameters fall into the regions of
       the phase-space density enhancements at $(V_\mathrm{C},
       \mathrm{FWHM}) \approx (-131_{-43}^{+33}, 27_{-7}^{+9}),
       (164_{-49}^{+71}, 26_{-7}^{+9})$ and
       $(-49_{-14}^{+11}~\mathrm{km\,s}^{-1},
       23_{-7}^{+10}~\mathrm{km\,s}^{-1})$, where the indexes indicate
       the half widths at the level of half maximum (HWHM) of the
       enhancements. The sky distribution of the Gaussians,
       corresponding to the first two concentrations, very well
       represents the sky distribution of HVCs, as obtained on the basis
       of the traditional definition of these objects. The Gaussians of
       the last concentration correspond to IVCs. Based on this
       identification, the division line between IVCs and HVCs can be
       drawn at about $|V_\mathrm{C}| = 74~\mathrm{km\,s}^{-1}$, and
       IVCs can be identified down to velocities of about
       $|V_\mathrm{C}| = 24~\mathrm{km\,s}^{-1}$. Traces of both IVCs
       and HVCs can also be seen in the sky distribution of Gaussians
       with $\mathrm{FWHM} \approx 7.3~\mathrm{km\,s}^{-1}$. In HVCs,
       these cold cores have small angular dimensions and low observed
       brightness temperatures $T_\mathrm{b}$. In IVCs, the cores are
       both larger and brighter.}
      {When neglecting the general decrease in the amount of gas at
       higher $|V_\mathrm{LSR}|$, the IVCs and HVCs are observed as
       distinctive maxima in the $(V_\mathrm{C}, \mathrm{FWHM})$
       distribution of the Gaussians, representing the structure of the
       21 cm radio lines of the Galactic \ion{H}{i}. This definition is
       less dependant than the traditional one, on the differential
       rotation model of the Galaxy. The consideration of line-width
       information may enable IVCs and HVCs to be better distinguished
       from each other, and from the ordinary Galactic \ion{H}{i}.}


   \maketitle

   \section{Introduction}

      In historical reviews, Wakker (\cite{Wak04}) and Wakker et al
      (\cite{Wea04}) state that early surveys of high-velocity
      \ion{H}{i} clouds (HVCs), completed in the Netherlands, were
      concentrated at high Galactic latitudes, where normal disc gas has
      a low velocity. Velocities were reported relative to the Local
      Standard of Rest (LSR), using a limit of $|V_\mathrm{LSR}| > 90$
      or $100~\mathrm{km\,s}^{-1}$ to define HVCs. During the course of
      these surveys, gas at velocities between $-50$ and
      $-100~\mathrm{km\,s}^{-1}$ was also found in many different
      directions. These objects were called intermediate-velocity clouds
      (IVCs). In this situation, a division of tasks was decided upon:
      astronomers of Leiden would analyse the high-velocity gas
      ($V_\mathrm{LSR} < -80~\mathrm{km\,s}^{-1}$), while those at
      Groningen would concentrate on the IVCs.

      As we can see, the velocity limits in these definitions are
      arbitrary, uncertain and partly based on historical arguments. A
      more physical approach to distinguishing of HVCs was introduced by
      Wakker (\cite{Wak91}), who defined the ``deviation velocity'',
      $V_\mathrm{DEV}$, to be the difference between the LSR velocity of
      the cloud and the extreme velocity allowed by Galactic
      differential rotation in a particular direction. However, de Heij
      et al. (\cite{Hei02}) presented a slightly different definition of
      the deviation velocity, which allows for a Galactic Warp.

      The definition of IVCs has remained uncertain. Albert \& Danly
      (\cite{Alb04}) state that IVCs are dynamically significant gas
      with velocities outside the range of the sum of Galactic rotation
      and Galactic velocity dispersion, but not as extreme as the
      velocities of HVCs. Numerically, they propose the velocity range
      $20 < |V_\mathrm{LSR}| < 100~\mathrm{km\,s}^{-1}$, but velocities
      below $40~\mathrm{km\,s}^{-1}$ can often be caused by differential
      Galactic rotation. Some authors have argued that the division
      between HVCs and IVCs, based on the LSR velocity, may be
      artificial. However, Wakker (\cite{Wak01}) found that IVCs appear
      to exist approximately $1~\mathrm{kpc}$ from the Galactic plane,
      while HVCs with known distance limits lie at $|z| >
      3~\mathrm{kpc}$ (Wakker \cite{Wak04}).

      \begin{figure*}
      \sidecaption
         \includegraphics[width=12cm]{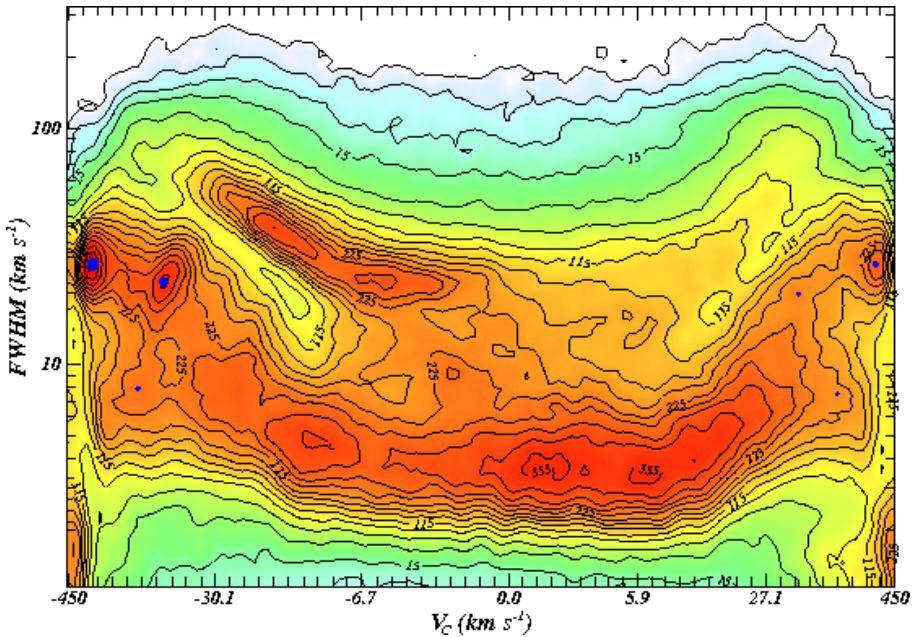}
         \caption{Distribution of Gaussian parameters in the
            $(V_\mathrm{C}, \mathrm{FWHM})$ plane. The x-axis is linear
            for the queue numbers of the Gaussians in the ascending
            sequence of the component velocities. The corresponding
            velocity scale is stretched around $V_\mathrm{C} =
            0~\mathrm{km\,s}^{-1}$ and compressed at higher velocities.
            To illustrate this, the ticks on the x-axis are drawn at
            $-450$, $-184$, $-117$, $-94$, $-75$, $-63$, $-54$, $-47$,
            $-41$, $-35$, $-30.1$, $-25.6$, $-21.8$, $-18.5$, $-15.6$,
            $-13.3$, $-11.4$, $-9.9$, $-8.6$, $-7.6$, $-6.7$, $-5.8$,
            $-5.0$, $-4.3$, $-3.6$, $-3.00$, $-2.38$, $-1.78$, $-1.18$,
            $-0.57$, 0.00, 0.56, 1.14, 1.71, 2.29, 2.89, 3.6, 4.3, 5.1,
            5.9, 6.9, 7.9, 9.2, 10.7, 12.5, 14.9, 17.9, 21.9, 27.1, 34,
            43, 54, 65, 79, 95, 121, 206 and $450~\mathrm{km\,s}^{-1}$.
            The y-axis is logarithmic in FWHM. The contour lines are
            drawn at the levels of 1, 3, 7, 15, 35, 60, 90, 115, 145,
            170, 195, 225, 250, 280, 315, 355 395 and 445 Gaussians per
            counting bin. The frequency enhancements, discussed in the
            text, are marked with blue circles.}
         \label{Fig01}
      \end{figure*}

      The situation with HVCs and IVCs is further complicated by the
      fact that the gas with a velocity, which is dynamically unusual or
      significant in its local or Galactic frame, may appear close to
      $0~\mathrm{km\,s}^{-1}$ with respect to the LSR. Therefore, it may
      be interesting to define these objects in way that is not based
      only on models of Galactic differential rotation. In this paper we
      try to identify both IVCs and HVCs using the frequency
      distribution of the Gaussian component parameters, derived in the
      decomposition of the Leiden/Argentine/Bonn (LAB) full sky database
      of \ion{H}{i} profiles (Kalberla et al \cite{Kal05}).

      A detailed justification for the use of Gaussian decomposition, in
      such a study, was provided by Haud \& Kalberla (\cite{Hau07},
      hereafter Paper III). The program that decomposes data from large
      21 cm \ion{H}{i} line surveys into Gaussian components, was
      described in the first paper of this series (Haud \cite{Hau00},
      hereafter Paper~I). The observational data for decomposition are
      from the LAB database of \ion{H}{i} 21 cm line profiles, which
      combines the new revision (LDS2, Kalberla et al. \cite{Kal05}) of
      the Leiden/Dwingeloo Survey (LDS, Hartmann \cite{Har94}), and a
      similar Southern sky survey (IARS, Bajaja et al. \cite{Baj05})
      completed at the Instituto Argentino de Radioastronomia. The LAB
      database is described in detail by Kalberla et al (\cite{Kal05}). 
      Our method of Gaussian decomposition generated 1\,064\,808
      Gaussians for 138\,830 profiles from LDS2, and 444\,573 Gaussians
      for 50\,980 profiles from IARS.
    
      In Paper~II (Haud \& Kalberla \cite{Hau06}), we analysed the
      distributions of the parameters of the obtained Gaussians. We
      focused mainly on the separation of the components describing
      different artefacts of the observations (interferences), reduction
      (baseline problems) and the decomposition (separation of signal
      from noise) process. In Paper~III, we introduced the study of the
      width distributions of the Gaussian components at
      equally-populated velocity intervals, and demonstrated that for
      Gaussians with relatively small LSR velocities ($-9 \le
      V_\mathrm{C} \le 4~\mathrm{km\,s}^{-1}$) it is possible to
      distinguish three or four groups of preferred line widths. The
      mean widths of these groups are $\mathrm{FWHM} = 3.9 \pm 0.6$,
      $11.8 \pm 0.5$, $24.1 \pm 0.6$, and $42 \pm
      5~\mathrm{km\,s}^{-1}$. In the present paper, we continue the
      analysis of the distribution of the Gaussian parameters, but focus
      on the components with the central LSR velocities $|V_\mathrm{C}|
      \ga 10~\mathrm{km\,s}^{-1}$. We demonstrate that some frequency
      enhancements in this region define two classes of objects whose
      general properties are similar to those of the IVCs and HVCs,
      defined by more traditional velocity criteria.

   \section{The velocity -- width distribution of the Gaussians}

      In Paper~III, we introduced the diagram (Fig. 2. of Paper III) of
      the frequency distributions of Gaussian component widths at
      equally-populated velocity intervals. We define the Gaussians by
      the standard formula
      \begin{equation}
         T_\mathrm{b} = T_\mathrm{b0}
                   \mathrm{e}^{-\frac{(V - V_\mathrm{C})^2}
                   {2\sigma_V^2}}, \label{Eq1}
      \end{equation}
      where $T_\mathrm{b}$ is the brightness temperature, and $V$ is the
      velocity of the gas relative to the Local Standard of Rest. 
      $T_\mathrm{b0} > 0$ is the height of the Gaussian at its central
      velocity $V_\mathrm{C}$. We characterise the widths of the
      components by their full width at the level of half maximum
      (FWHM), which is related to the velocity dispersion $\sigma_V$,
      obtained from our decomposition program, by a simple scaling
      relation $\mathrm{FWHM} = \sqrt{8 \ln{2}} \sigma_V$.

      Because the Gaussian functions fitted to the complex \ion{H}{i}
      profiles, close to the Galactic plane, cannot be directly
      interpreted in terms of the properties of gas clouds, in Paper III
      we focused on the higher Galactic latitudes, and stressed the
      differences in the $(V_\mathrm{C}, \mathrm{FWHM})$ frequency
      distribution of Gaussians at different latitudes. In this paper,
      we concentrate on higher velocities, where differences in the
      distributions for different latitude ranges are smaller, and
      therefore present, in Fig.~\ref{Fig01}, the distribution of
      Gaussians for all Galactic latitudes. To construct this figure, we
      arranged all Gaussians with central velocities in the
      decomposition range ($-460 < V_\mathrm{C} <
      396~\mathrm{km\,s}^{-1}$ in the LDS2 and $-437 < V_\mathrm{C} <
      451~\mathrm{km\,s}^{-1}$ in the IARS; for details see Papers~I and
      II), in ascending order of their $V_\mathrm{C}$. We then grouped
      the sequence into 129 bins of an equal number of Gaussians,
      rejecting some Gaussians with the most extreme velocities. We
      binned the line widths in equal steps of $\lg(\sigma_V)$ of 0.025.
      The isolines in Fig.~\ref{Fig01} provide the number of Gaussians
      at each of such two-dimensional parameter interval.

      In Fig.~\ref{Fig01}, the region most densely populated by
      Gaussians, lies in the lower part of the figure at relatively
      small LSR velocities. In Paper III, we demonstrated that most of
      these Gaussians represent the cold neutral medium of the Galactic
      disc. In the present figure, this low-velocity region is also
      seriously contaminated by Gaussians from the strongest \ion{H}{i}
      profiles, close to the Galactic plane, parameters of which are not
      directly related to the properties of the real gas. At somewhat
      higher negative velocities (from $V_\mathrm{C} \approx -2$ to
      $-40~\mathrm{km\,s}^{-1}$) and larger line widths (from
      $\mathrm{FWHM} \approx 20$ to $70~\mathrm{km\,s}^{-1}$), we can
      see an elongated region of high Gaussian density. We will not
      discuss this feature in the present paper. At even higher negative
      velocities in Fig.~\ref{Fig01}, we can see two high-density
      regions centred at $(V_\mathrm{C}, \mathrm{FWHM}) \approx
      (-131_{-43}^{+33}, 27_{-7}^{+9})$ and
      $(-49_{-14}^{+11}~\mathrm{km\,s}^{-1},
      23_{-7}^{+10}~\mathrm{km\,s}^{-1})$ (the largest blue circles in
      Fig.~\ref{Fig01}). Here the indexes of the values of the
      velocities and line widths provide estimates of the half widths of
      the corresponding distributions at the level of half maximum. At
      positive velocities, only one such frequency enhancement is
      clearly visible about $(V_\mathrm{C}, \mathrm{FWHM}) \approx
      (164_{-49}^{+71}~\mathrm{km\,s}^{-1},
      26_{-7}^{+9}~\mathrm{km\,s}^{-1})$. These enhancements contain the
      Gaussians of our main interest in the present paper. At even
      higher negative and positive velocities, two concentrations of
      narrow Gaussians are clearly visible (at the bottom corners of the
      figure). These are generated by the presence of spurious
      Gaussians, as discussed in Paper II.

      In Paper III, we demonstrated that, for Gaussians with relatively
      small LSR velocities, it is possible to distinguish three or four
      groups of preferred line widths. Similar line-width groups appear
      to exist at higher velocities, but their mean widths depend on the
      velocity of the selected Gaussians, and only in relatively small
      velocity intervals, we can consider the line-width distributions
      to be nearly independent of the central velocity of the Gaussians.
      Therefore, we must require a separate model of the width
      distribution, for each equally-populated velocity interval. Only
      this would make it possible to distinguish between components,
      belonging to different line-width groups.
    
      As in the case of lower velocities, we model all the width
      distributions with a sum of lognormal functions defined by
      \begin{equation}
         f_{\mu, \sigma^2} (x) = \left\{
         \begin{array}{ll}
            0 & (x \le 0)\\
            \frac{f_0}{\sqrt{2\pi}\sigma x}
            \mathrm{e}^{-\frac{(\ln(x) - \mu)^2}
            {2\sigma^2}} & (x > 0)\,,
         \end{array}\right. \label{Eq2}
      \end{equation}
      where $f_0$, $\mu$ and $\sigma$ may be considered as free
      parameters for fitting the model distribution to the observed one.
      As in Paper III, we used, in most cases, four lognormal functions
      for each modelled velocity interval. Only at the highest positive
      and negative velocities, we introduced an additional lognormal to
      reject the spurious Gaussians, causing the enhancements at the
      lower corners of Fig.~\ref{Fig01}.

      Many different density functions are available that can model the
      distributions of Gaussian widths, which are defined to be
      positive. We chose to use the lognormal function because its usage
      is technically convenient: if a parameter is distributed according
      to the lognormal law, the distribution of the logarithms of the
      parameter is described by a Gaussian. Mebold (\cite{Meb72}) also
      argued that his narrow and shallow components could be identified
      most easily, when the line-width distribution is studied in a
      logarithmic scale. Moreover, from all tested density functions,
      the lognormal provided the best fits.

\begin{figure}
         \resizebox{\hsize}{!}{\includegraphics{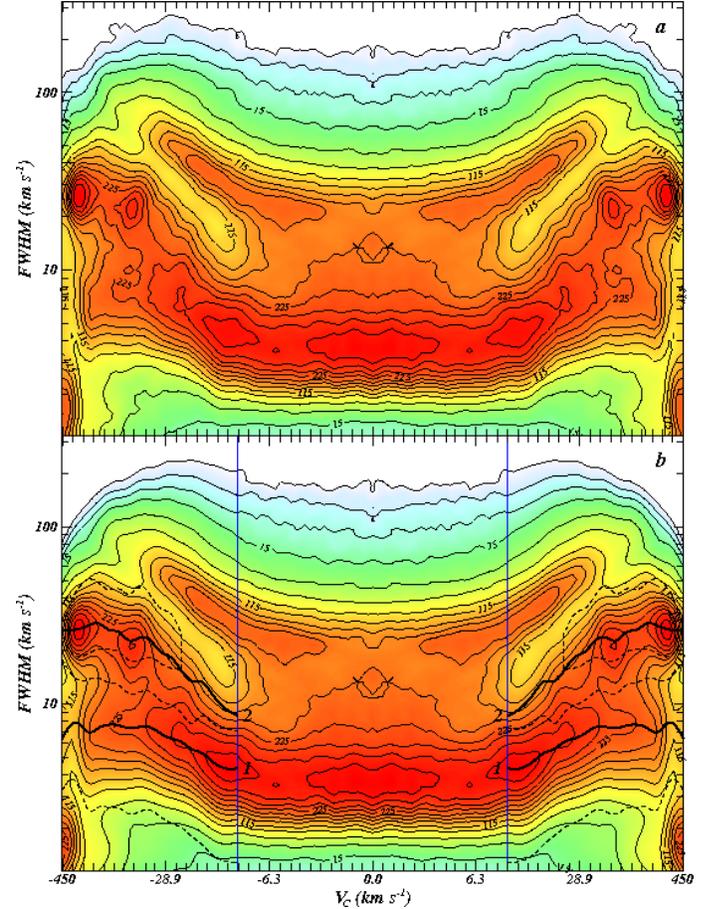}}
         \caption{Symmetrised distribution of Gaussian parameters in the
            $(V_\mathrm{C}, \mathrm{FWHM})$ plane (a) and the
            corresponding model distribution (b). The contour lines are
            obtained in the same way as in Fig.~\ref{Fig01}. The ticks
            on the x-axis are drawn at 0.00, $\pm 0.54$, $\pm 1.10$,
            $\pm 1.66$, $\pm 2.22$, $\pm 2.79$, $\pm 3.4$, $\pm 4.1$,
            $\pm 4.7$, $\pm 5.5$, $\pm 6.3$, $\pm 7.2$, $\pm 8.2$, $\pm
            9.3$, $\pm 10.8$, $\pm 12.4$, $\pm 14.5$, $\pm 17.1$, $\pm
            20.3$, $\pm 24.2$, $\pm 28.9$, $\pm 35$, $\pm 41$, $\pm 49$,
            $\pm 57$, $\pm 67$, $\pm 80$, $\pm 98$, $\pm 122$, $\pm 203$
            and $\pm 450~\mathrm{km\,s}^{-1}$. The vertical blue lines
            indicate the range of modelling $(|V_\mathrm{C}| \ge
            9.38~\mathrm{km\,s}^{-1})$, the thick black lines mark the
            run of the peaks of the two line-width groups and the dashed
            lines on both sides of the thick lines give the extent of
            the regions, where the Gaussians belong respectively to
            group 1 or 2 with the probability higher than 75\%.}
         \label{Fig02}
      \end{figure}

      The described modelling was relatively easy for the negative
      velocity part of the distribution in Fig.~\ref{Fig01}, but more
      ambiguous for positive velocities, as there the number of
      Gaussians is smaller, equally-populated velocity intervals are
      wider, and the density maxima are less obvious. At the same time,
      it can be seen from Fig.~\ref{Fig01} that the general structure of
      the distribution is similar for negative and positive velocities. 
      The enhancements are weaker at positive velocities, but we may
      find that these weaker maxima are at approximately the same
      positions as for negative velocities: a strong enhancement at
      $(V_\mathrm{C}, \mathrm{FWHM}) \approx (-131~\mathrm{km\,s}^{-1},
      27~\mathrm{km\,s}^{-1})$ corresponds to a weaker one at
      $(164~\mathrm{km\,s}^{-1}, 26~\mathrm{km\,s}^{-1})$, and a weaker
      enhancement at $(-49~\mathrm{km\,s}^{-1}, 23~\mathrm{km\,s}^{-1})$
      corresponds to an even weaker one at about
      $(44~\mathrm{km\,s}^{-1}, 20~\mathrm{km\,s}^{-1})$. Therefore, to
      consider the positive velocity information, we decided to average
      the positive and negative velocity sides of the distribution
      (Fig.~\ref{Fig02}a), and to construct the same model for the
      positive and negative velocities (Fig.~\ref{Fig02}b).

      \begin{figure}
         \resizebox{\hsize}{!}{\includegraphics{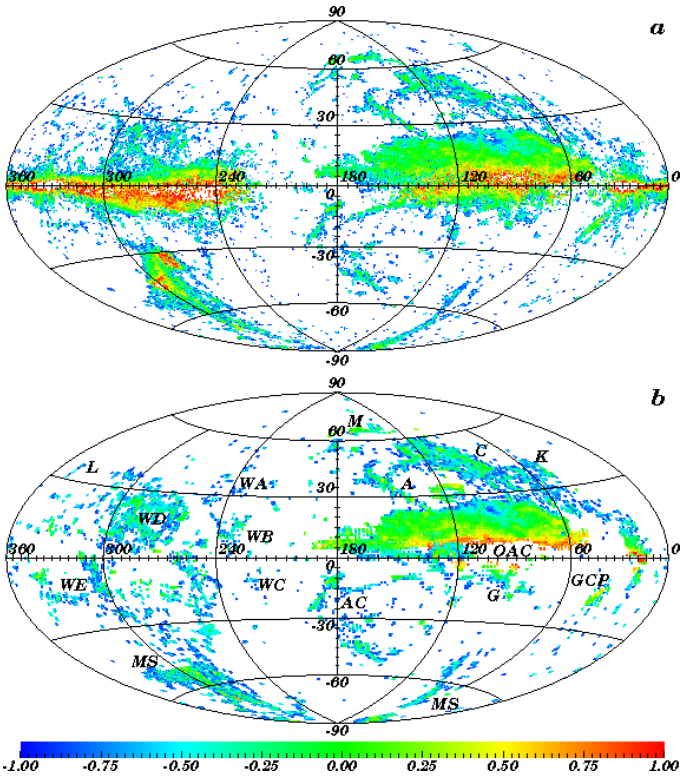}}
         \caption{Sky distribution in Galactic coordinates of the
            Gaussians, belonging to the second line-width group at
            velocities $|V_\mathrm{C}| > 74~\mathrm{km\,s}^{-1}$ (a) and
            the brightness of the high-velocity hydrogen clouds from the
            Hulsbosch \& Wakker (\cite{Hul88}) and Morras et al.
            (\cite{Mor00}) surveys (b). The labels with capital letters
            indicate the larger HVC complexes according to Wakker
            (\cite{Wak04}). In the upper panel the colours encode the
            values of $\lg (T_p)$. In the lower panel the same colour
            scale corresponds to the logarithms of the brightness
            temperatures of the HVC detections. In both panels near the
            Galactic plane the strongest Gaussians with $T_p$ or
            $T_\mathrm{b0} > 10~\mathrm{K}$ are not plotted.}
         \label{Fig03}
      \end{figure}

      \begin{figure}
         \resizebox{\hsize}{!}{\includegraphics{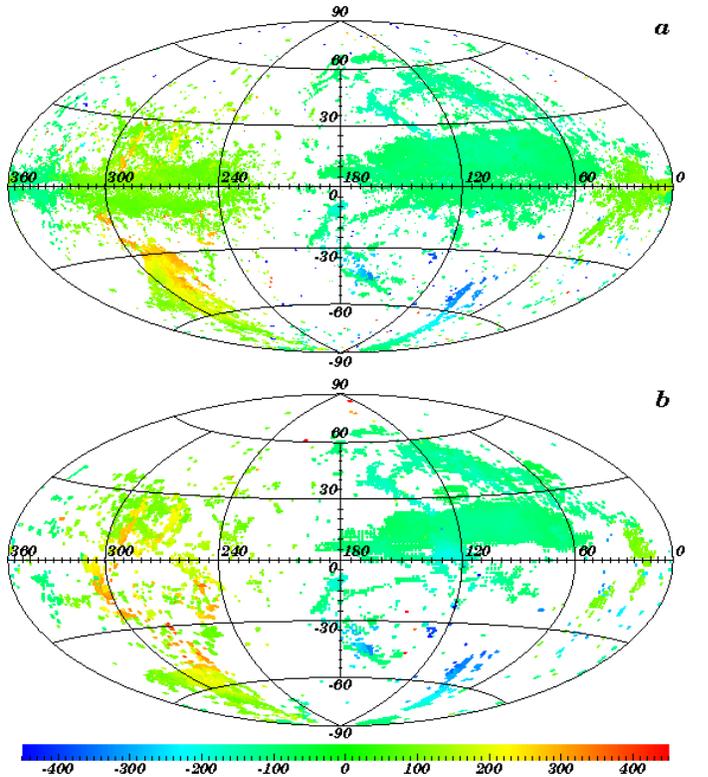}}
         \caption{Same as Fig.~\ref{Fig03}, but the colours represent
            the LSR velocities of the gas.}
         \label{Fig04}
      \end{figure}

      During the modelling, we encountered one additional problem. When
      the values of $f_0$ and $\mu$ for lognormal distributions were
      well determined, the values of $\sigma$ of different lognormal
      functions were often strongly correlated with each other. This
      occasionally meant that the $\sigma$ of the same line-width
      component, had very different values for neighbouring velocity
      intervals. Numerically, when the mean uncertainties in fitting
      $f_0$ and $\mu$ were 8.6\% and 2.5\%, respectively, the
      corresponding uncertainty in the values of $\sigma$ was about
      38\%. To overcome this problem, we first fitted all width
      distributions, considering $f_0$, $\mu$ and $\sigma$ of all
      lognormal functions as free parameters. We then analysed the
      $\sigma$ of each line-width component, as a function of the number
      of the velocity interval, and applied natural smoothing splines. 
      In the second run of fitting of the sums of lognormal functions to
      the width distributions of Gaussians in different
      equally-populated velocity intervals, we replaced the values of
      $\sigma$, obtained in the first fitting, with the results of the
      smoothing, and kept the values of $\sigma$ fixed while iterating
      $f_0$ and $\mu$. The resulting model distribution is provided in
      Fig.~\ref{Fig02}b. We are not interested in velocities close to
      zero, which were discussed in Paper III. The modelling was
      performed only for $|V_\mathrm{C}| \ge 9.38~\mathrm{km\,s}^{-1}$,
      indicated in Fig.~\ref{Fig02}b with blue vertical lines.

      After defining line-width groups of Gaussians using different
      lognormal functions, it is interesting to find out how these
      groups are related to actual gas within and about the Galaxy. To
      study this question, we need to determine which Gaussian belongs
      to which line-width group. The frequency distributions of
      Gaussians of different groups, however, overlap considerably and
      we are unable to identify with full confidence the Gaussians
      belonging to any particular group. However, using our model
      distribution, we can estimate for every Gaussian the probability
      of belonging to some particular line-width group.

      So, let us define the probability that a Gaussian belongs to the
      line-width group $i$ as
      \begin{equation}
         p_i = \frac{f_i (V_\mathrm{C}, \sigma_V)}
              {\sum_{j=1}^n f_j (V_\mathrm{C}, \sigma_V)}\,, \label{Eq3}
      \end{equation}
      where $f_i$ is the density function of the lognormal distribution,
      as defined by Eq.~\ref{Eq2}, corresponding to the $i$-th
      line-width group, $V_\mathrm{C}$ is the central velocity, and
      $\sigma_V$ the width of the particular Gaussian and $n = 4$ or 5,
      depending on the velocity interval under consideration. To study
      the sky distribution of the Gaussians from the selected line-width
      group $i$, we can then plot $T_p = T_\mathrm{b,0} p_i$, the
      heights of Gaussians $T_\mathrm{b,0}$, multiplied by the
      probability that the Gaussian belongs to this line-width group
      $p_i$. In this way, all the Gaussians from our decomposition will
      be presented on each plot. However, the brightness of the
      Gaussians, for which the probability of belonging to a particular
      line-width group is low, is suppressed considerably by these small
      probabilities. To enhance the contrast in plots even more, and
      because the noise level of the LAB Survey is about $\langle
      \sigma_\mathrm{rms}\rangle = 0.09$~K (see Paper II), we also do
      not plot Gaussians for which $T_p < 0.1~K$. In this way we
      exclude, from the sky plots, most of the weak Gaussians, which
      represent the survey baseline problems. Stronger spurious
      Gaussians are excluded, according to the criteria described in
      Paper II.

   \section{High velocities}

      As stated in the previous section, we are mostly interested in
      frequency enhancements in Fig.~\ref{Fig01}, which are centred at
      $(V_\mathrm{C}, \mathrm{FWHM}) \approx (-131, 27)$, $(-49, 23)$,
      $(44, 20)$ and $(164~\mathrm{km\,s}^{-1},
      26~\mathrm{km\,s}^{-1})$. In our model, these enhancements
      correspond to the line-width group 2 as labelled in
      Fig.~\ref{Fig02}b. As can be seen from Fig.~\ref{Fig01}, at
      negative velocities this line-width group clearly forms two
      frequency enhancements in different velocity regions. These
      frequency enhancements are separated by the region, more dominated
      by Gaussians of other line-width groups, and the second group has
      a frequency minimum at about $|V_\mathrm{C}| =
      74_{-6}^{+7}~\mathrm{km\,s}^{-1}$. The same appears to be true for
      positive velocities, but there the picture is considerably less
      clear.

      We concentrate first on Gaussians of the second line-width group
      with $|V_\mathrm{C}| > 74~\mathrm{km\,s}^{-1}$. Fig.~\ref{Fig03}a
      illustrates the sky distribution of brightness temperatures of
      these components, and Fig.~\ref{Fig04}a provides the distribution
      of LSR velocities for the same Gaussians. In Fig.~\ref{Fig03}b, we
      present the sky distribution of brightness temperatures of HVCs
      with $T_\mathrm{b} > 0.1~K$ as compiled from Hulsbosch \& Wakker
      (\cite{Hul88}) and Morras et al. (\cite{Mor00}) catalogues, and
      Fig.~\ref{Fig04}b provides their velocity distribution. Comparing
      the panels of these figures, a surprising similarity is visible
      between the distributions for the objects, selected from different
      observational data by different procedures.

      The HVC catalogues by Hulsbosch \& Wakker (\cite{Hul88}) and
      Morras et al. (\cite{Mor00}) are based on \ion{H}{i} profiles with
      a velocity resolution of about $16~\mathrm{km\,s}^{-1}$ and a
      noise level of about $\langle \sigma_\mathrm{rms}\rangle =
      0.015$~K. All the observed profiles were scanned visually for the
      presence of components at high velocities, using two selection
      criteria: the component should have a brightness temperature
      $T_\mathrm{b} \ge 0.05-0.08~K$ and should have $|V_\mathrm{LSR}|
      \ge 100~\mathrm{km\,s}^{-1}$ or $80~\mathrm{km\,s}^{-1}$. 
      Figs.~\ref{Fig03}a and \ref{Fig04}a are based on the LAB Survey,
      which has a velocity resolution of about $1~\mathrm{km\,s}^{-1}$,
      and a noise level of about 0.09~K. Both figures were made without
      any visual inspection of the profiles, using statistical data for
      Gaussians (mostly $T_\mathrm{b} \ge 0.13~K$) obtained in a fully
      automatic decomposition of the survey profiles. In our study the
      first selection criterion was the line width, which was not
      directly used by Hulsbosch \& Wakker (\cite{Hul88}) and Morras et
      al. (\cite{Mor00}). Nevertheless, all well-known HVC complexes are
      clearly visible in both figures.

\begin{figure}
         \resizebox{\hsize}{!}{\includegraphics{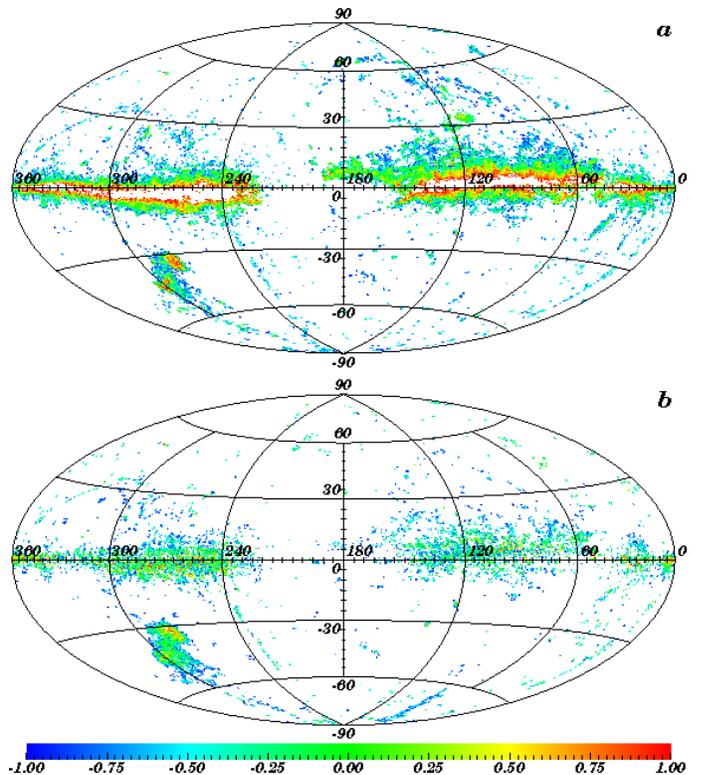}}
         \caption{Same as Fig.~\ref{Fig03}a, but for the line-width
            group 1 (a) and for all Gaussians, except those belonging to
            the line-width groups 1 and 2 (b).}
         \label{Fig05}
      \end{figure}

      Besides the differences, described above, there is a selection
      criterion almost in common for panels a and b of Figs.~\ref{Fig03}
      and \ref{Fig04} -- the velocity range. For panels a of these
      figures, we have used Gaussians with $|V_\mathrm{C}| >
      74~\mathrm{km\,s}^{-1}$. Figs.~\ref{Fig03}b and \ref{Fig04}b
      represent the gas with $|V_\mathrm{LSR}| \ge
      100~\mathrm{km\,s}^{-1}$ or $80~\mathrm{km\,s}^{-1}$. It is
      possible that the similarity of these velocity limits is the main
      factor that determines the similarity of the sky distributions. 
      However, in this case, Figs.~\ref{Fig03}a and \ref{Fig04}a must
      remain unchanged if we keep the velocity limits, defined above,
      but instead of Gaussians from line-width group 2, use those from
      some other group (number 1, for example). The actual results do
      not confirm this expectation. The sky distribution of the
      Gaussians of the first line-width group (Fig.~\ref{Fig05}a) is
      different from that of the second line-width group
      (Fig.~\ref{Fig03}a). The well-populated HVC complexes are no
      longer visible. In Fig.~\ref{Fig05}a, only small concentrations of
      weak dots are observed in the regions, which in Fig.~\ref{Fig03}a
      are populated by the largest HVC complexes.

      This result becomes understandable in the light of the discussion
      by Kalberla \& Haud (\cite{Kal06}): most of the high-velocity
      clouds have a well-defined two-component structure, where the cold
      HVC phase has a typical line-width of about
      $7.3~\mathrm{km\,s}^{-1}$, and exists only within more extended
      broad-line regions, typically with $\mathrm{FWHM} \approx
      27.2~\mathrm{km\,s}^{-1}$. Therefore, we may state that the
      modelling of the line-width distribution of Gaussians permits us
      to separate the warm and cold gas phases in HVCs, as it was also
      possible for the local \ion{H}{i} gas. However, for the local gas
      the corresponding line widths were $\mathrm{FWHM} = 3.9$ and
      $24.1~\mathrm{km\,s}^{-1}$.

      There are still Gaussians belonging neither to line-width
      component 1 nor to component 2. Their sky distribution is given in
      Fig.~\ref{Fig05}b. This figure shares the same velocity limit as
      Figs.~\ref{Fig03}a and \ref{Fig05}a, but here we can see only
      rather weak traces of all HVC structures. Of course, it would be
      nice, if in this figure there were no traces of HVCs at all, but
      we must remember that the separation of the Gaussians into
      different line-width groups is only statistical and the
      probabilities are determined from the smoothed model. It seems
      that this model has worked most badly for some parts of the
      Magellanic Stream.

      Besides the similarities between Figs.~\ref{Fig03}a and
      \ref{Fig03}b, there are also considerable differences. The most
      obvious one is the presence in Fig.~\ref{Fig03}a of a large number
      of strong Gaussians, close to the Galactic plane and in the region
      of the Magellanic Clouds. However, close to the Galactic plane the
      Gaussians are obtained from the decomposition of very complex
      \ion{H}{i} profiles of the Galactic disc, in which the real gas
      structures are heavily blended with each other; these Gaussians
      should therefore not be considered as directly representing the
      properties of the ISM. The same regions are heavily populated in
      both panels of Fig.~\ref{Fig05}. In these regions, the
      decomposition provides the Gaussians of all possible parameters,
      and without any concentration of these parameter values in any
      region of parameter space. In Fig.~\ref{Fig01}, these Gaussians
      form the general background against which we may distinguish the
      features, more directly representing the properties of Galactic
      \ion{H}{i}. The same holds also for the main bodies of the
      Magellanic Clouds.

      \begin{figure}
         \resizebox{\hsize}{!}{\includegraphics{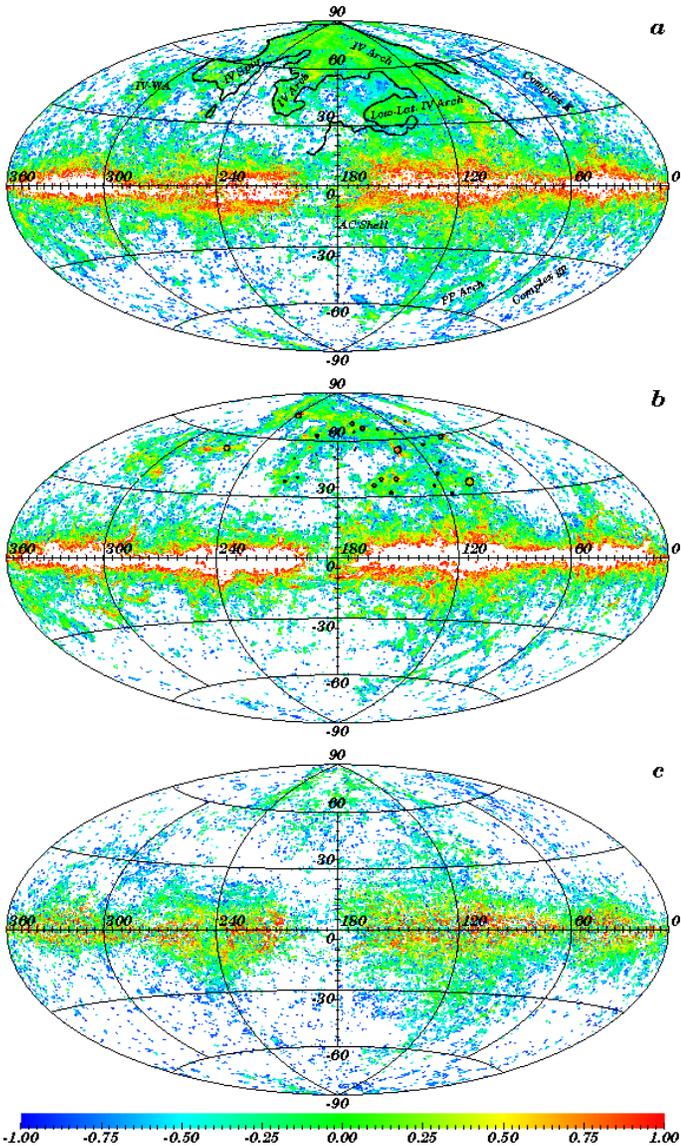}}
         \caption{Same as Fig.~\ref{Fig03}a, \ref{Fig05}a and
            \ref{Fig05}b, but for the velocity interval $24 <
            |V_\mathrm{C}| < 74~\mathrm{km\,s}^{-1}$. (a) The sky
            distribution of the Gaussians, belonging to the second
            line-width group. The solid contour and the labels indicate
            the region of major intermediate-velocity gas features
            according to Kuntz \& Danly (\cite{Kun96}) and Wakker
            (\cite{Wak01}). (b) The sky distribution of the Gaussians
            belonging to the first line-width group. The circles
            indicate the locations of the intermediate-velocity gas
            clumps from the catalogue by Kuntz \& Danly (\cite{Kun96}).
            (c) The sky distribution of all Gaussians, except the ones
            corresponding to the first and second line-width group.}
         \label{Fig06}
      \end{figure}

      \begin{figure}
         \resizebox{\hsize}{!}{\includegraphics{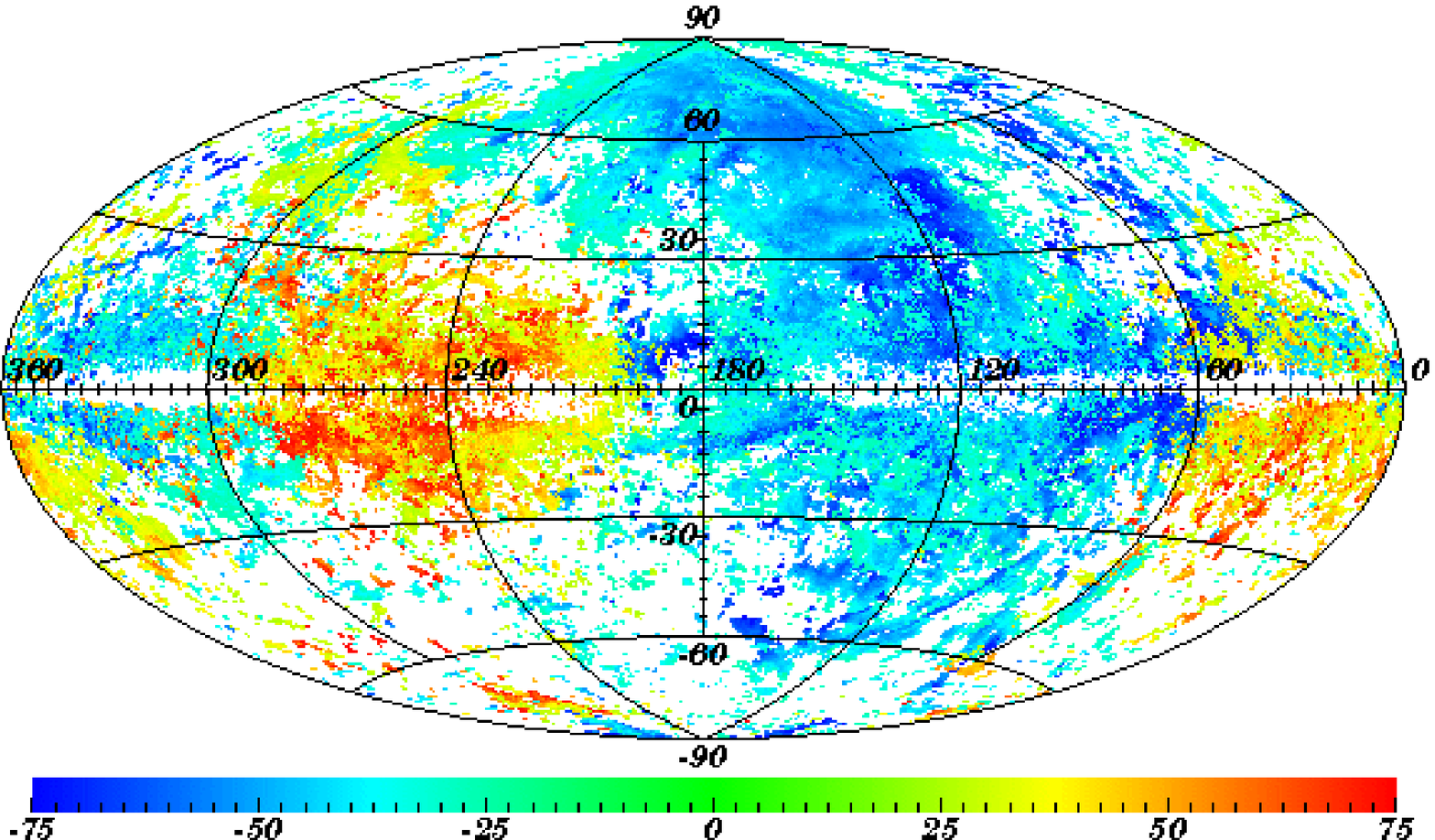}}
         \caption{Same as Fig.~\ref{Fig06}a, but the colours represent
            the LSR velocities of the gas.}
         \label{Fig07}
      \end{figure}

      As we can estimate from the comparison of different sky
      distribution figures, close to the Galactic plane, the region of
      ``confusion'' lies generally at $|b| < 15\degr$. Above this
      latitude limit, the distribution of Gaussians appears to be
      representative of the properties of HVCs. This is of particular
      importance for the Outer Arm Cloud (OAC). In early studies (Habing
      \cite{Hab66}, Hulsbosch \& Wakker \cite{Hul88}), this complex was
      considered to be an HVC. Verschuur (\cite{Ver75}) and Haud
      (\cite{Hau92}) have modelled the complex as part of the outer disc
      of the Milky Way. Based on the deviation velocity, a recent review
      (Wakker \cite{Wak04}) considers the OAC to be an
      intermediate-velocity cloud. From a statistical analysis of the
      Gaussian parameters, we are unable to conclude much about the OAC
      close to the Galactic plane. In the parts most distant from the
      galactic plane however, the gas properties appear to agree with
      the gas properties in HVCs.
    
      In conclusion, we note that two different approaches to
      identifying high-velocity \ion{H}{i} gas at high Galactic
      latitudes ($|b| > 15\degr$), have produced remarkably similar
      results. We cannot, of course, consider these methods to be
      completely independent ones. The classical definition of HVCs is
      fully based on their LSR velocities, and the model of differential
      rotation of the Galactic disc. The horizontal structure of
      Fig.~\ref{Fig01} is largely determined by the Galactic
      differential rotation as well, but a comparison of
      Figs.~\ref{Fig03}a, \ref{Fig05}a and \ref{Fig05}b demonstrates
      that velocity is not the only factor that defines HVCs. Moreover,
      when using the statistical distribution of Gaussian parameters, we
      can estimate the extent of the velocity and line-width intervals
      populated by the components, which most likely represent HVCs. 
      Unfortunately, the analysis becomes complicated and unreliable,
      when approaching the Galactic plane, where the classic definition
      of HVCs also has the greatest trouble.

   \section{Intermediate velocities}

      We have so far considered line-width groups only for velocities
      $|V_\mathrm{C}| > 74~\mathrm{km\,s}^{-1}$. There is however,
      another maximum at lower velocities, which peaks at $V_\mathrm{C}
      \approx -49~\mathrm{km\,s}^{-1}$. The lower velocity limit of this
      frequency enhancement is questionable. The concentration about
      $(V_\mathrm{C}, \mathrm{FWHM}) \approx (-49~\mathrm{km\,s}^{-1},
      23~\mathrm{km\,s}^{-1})$, in Fig.~\ref{Fig01}, reaches its HWHM at
      $V_\mathrm{C} \approx -38~\mathrm{km\,s}^{-1}$. The relative
      frequency of group-2 Gaussians drops to the level of the minimum
      between two peaks in Fig.~\ref{Fig02}b at about $|V_\mathrm{C}| =
      35~\mathrm{km\,s}^{-1}$, but even after this, group 2 can still
      modelled until about $|V_\mathrm{C}| = 24~\mathrm{km\,s}^{-1}$. At
      even lower velocities, the determined parameters of the second
      line-width group became rather uncertain, the corresponding
      Gaussians are relatively few in number, and we do not discuss
      these velocities.

      The sky distribution of the Gaussians of the second line-width
      group in the intermediate velocity range $24 < |V_\mathrm{C}| <
      74~\mathrm{km\,s}^{-1}$ is shown in Fig.~\ref{Fig06}a. Their
      velocity distribution is provided in Fig.~\ref{Fig07}. In
      Fig.~\ref{Fig06}a, we mark with a thick solid line, the $1.8
      \times 10^{19}~\mathrm{cm}^{-2}$ surface density contour of the
      intermediate-velocity hydrogen clouds, as published in Fig.~8 of
      Kuntz \& Danly (\cite{Kun96}). Some additional IVC complexes,
      identified by Wakker (\cite{Wak01}), are also labelled. We see
      that, as in the case of HVCs at higher velocities, in the
      intermediate velocity range the distribution of Gaussians, from
      the second line-width group, closely follows the main features of
      the distribution of the IVCs. This indicates that both HVCs and
      IVCs appear to represent related features in different velocity
      ranges.

      As for HVCs, in the intermediate velocity range, many Gaussians of
      the first line-width group are concentrated in sky regions
      populated by IVC gas, of the second line-width group
      (Fig.~\ref{Fig06}b). Among these are Gaussians, for which the
      probabilities of belonging to the first or second line-width group
      are nearly equal to each other. These Gaussians are plotted both
      in Figs.~\ref{Fig06}a and \ref{Fig06}b. A more detailed
      examination of the situation however, indicates that this is not
      the only effect. In many sky positions in the region covered by IV
      Arches and Spur, the decomposed hydrogen profiles contain
      Gaussians from both line-width groups. Altogether the profiles for
      36\,375 sky positions at $|b| \ge 30\degr$ contain Gaussians,
      which can be classified with at least 50\% confidence to belong to
      the first or second line-width group, and of these sky positions,
      9\,015 profiles, or nearly 25\%, contain both the Gaussians of the
      first and the second line-width group.
    
      The comparison of Figs.~\ref{Fig06}a and \ref{Fig06}b reveals
      differences in the properties of the IVC components, corresponding
      to the two different line-width groups. When the clouds
      corresponding to the first line-width group appear to be brighter
      and clumpier, those corresponding to the second line-width group
      cover the sky in the regions of IV Arches and Spur more uniformly,
      but with smaller average brightness. A similar behaviour by the
      intermediate-velocity gas was noted by Kuntz \& Danly
      (\cite{Kun96}), who stated that the bulk of the
      intermediate-velocity gas is primarily composed of clumps, which
      appear to be surrounded by a lower column density envelope.

      In their Table 1, Kuntz \& Danly (\cite{Kun96}) published the
      catalogue of denser clumps, which we have presented in
      Fig.~\ref{Fig06}b as circles centred at the positions of the peak
      column densities of the clumps. Unfortunately, the above-mentioned
      Table 1 does not contain much information about the shapes and
      sizes of the clumps. We therefore plot these objects in
      Fig.~\ref{Fig06}b, as circles of radii proportional to the clump
      peak intensities, which are provided by Kuntz \& Danly
      (\cite{Kun96}). As can be seen, the brighter clumps from the
      catalogue coincide with the concentrations of Gaussians in our
      figure. The correspondence is more problematic for the weakest
      clumps (for which the circles have been reduced in size to dots),
      but here we must consider that the paper by Kuntz \& Danly
      (\cite{Kun96}) was based on the Bell Laboratories \ion{H}{i}
      survey described by Stark et al., (\cite{Sta92}). The rms noise
      per channel for this survey was $0.017$~K, a value more than 5
      times lower than for LDS, such that many of the weakest clumps
      detected in these data may not be detected by us. Moreover, the
      sampling grid of the Bell Laboratories survey was $2\degr$ in
      declination and about $0\fdg5$ in right ascension, which was
      rebinned by Kuntz \& Danly (\cite{Kun96}) into a regular grid,
      comparable to the beam size of about $2\fdg5$, of the horn
      reflector used. As a result, the locations of the circles in
      Fig.~\ref{Fig06}b should not be considered to be accurate.

      Our results appear to confirm that the IVCs have a two-component
      structure: brighter clumps, whose emission lines correspond to the
      first line-width group; and lower column density envelopes, whose
      emission can be described by Gaussians of the second line-width
      group. A similar two-component structure is found for HVCs, for
      which the velocity profiles are often composed of a broad
      ($\mathrm{FWHM} \sim 20-25~\mathrm{km\,s}^{-1}$), and a narrow
      ($\sim 8~\mathrm{km\,s}^{-1}$) component (Wakker \& Woerden
      \cite{Wak97}; mean $\mathrm{FWHM} \approx
      27.2~\mathrm{km\,s}^{-1}$ and $7.3~\mathrm{km\,s}^{-1}$ according
      to Kalberla \& Haud \cite{Kal06}). The corresponding mean line
      widths for IVCs, are $\mathrm{FWHM} \approx
      22.2~\mathrm{km\,s}^{-1}$ and $7.2~\mathrm{km\,s}^{-1}$,
      respectively.

      Fig.~\ref{Fig06}c corresponds to Fig.~\ref{Fig05}b for the high
      velocity region and demonstrates that in the sky distribution of
      the Gaussians, not belonging to line-width groups 1 or 2, we can
      see only weak traces of the IVCs. These traces once again, are
      mostly caused by difficult-to-classify, relatively bright
      Gaussians, which appear both in Figs.~\ref{Fig06}a or \ref{Fig06}b
      and Fig.~\ref{Fig06}c. Moreover, the visibility of these traces is
      enhanced in Fig.~\ref{Fig06}c, because we do not plot in any sky
      distribution, the Gaussians, which are considered to be spurious,
      according to the selection criteria described in Paper II. By
      adding these components, Fig.~\ref{Fig06}c would be more
      homogeneously filled with random dots. Only the Galactic plane
      would remain visible, as a region of slightly enhanced
      concentration of the Gaussians. Close to the Galactic plane,
      however, the Gaussian decomposition itself is badly determined,
      and Gaussians of all possible parameters are found. This region is
      therefore, filled by dots in all sky distribution figures.

      Finally, according to our knowledge, so far there have not been
      any systematic searches of IVCs in the southern sky (the blank
      region in Fig. 3 of Wakker \cite{Wak04}). As LAB includes both the
      northern and the southern data, our Fig.~\ref{Fig06} also contains
      information on the southern IVCs. However, as we can see, in this
      figure the regions, observable only from the southern hemisphere
      of the Earth, are almost empty, and the only possible new IVC
      complexes are the concentrations of dots in the region $280\degr <
      l < 300\degr$, $-35\degr < b < -15\degr$. In this region, we can
      see 2-3 ``clouds'' with typical IVCs properties: relatively small
      and bright concentrations of the group-1 Gaussians, surrounded by
      wider envelopes of the weaker group-2 Gaussians.

   \section{Discussion}

      We have demonstrated that both the HVCs and IVCs could be
      identified as concentrations of Gaussians in $(V_\mathrm{C},
      \mathrm{FWHM})$ frequency plots. For both cloud types, we have
      identified a two-phase structure, where the cold phase,
      corresponding in Gaussian width distribution to group 1, exists
      within more extended broad-line (line-width group 2) regions. In
      this respect, both HVCs and IVCs appear to represent related
      dynamical features, in different velocity ranges. However, besides
      similarities, there are differences between the cloud types. The
      first difference can be seen in Fig.~\ref{Fig01}: at both negative
      and positive velocities the concentrations of Gaussians,
      corresponding to the envelopes of the IVCs, are located at lower
      line-widths than those corresponding to HVCs.

      We have noted differences in the relative intensities of the cores
      and envelopes of HVCs and IVCs. In HVCs, the components obtained
      from LAB, and corresponding to the cold cores, are relatively
      weak, and the angular dimensions of the cores found are relatively
      small. The cores in IVCs appear to be larger, with more intense
      radiation. This could imply that both the linear dimensions and
      gas content, of the cores of HVCs and IVCs, are similar, but IVCs
      are closer to us, on average, than the HVCs. The cores of the more
      distant HVCs would then have smaller angular sizes, and fill the
      aperture of a $25~\mathrm{m}$ telescope to a lesser extent than
      the more nearby cores of IVCs. As a result, they are detected more
      weakly. This interpretation is supported by Schwarz \& Wakker
      (\cite{Sch04}), who state that intermediate (10\arcmin) resolution
      observations of HVCs imply smaller core angular dimensions, and
      higher core brightnesses than measured at lower resolution. The
      explanation is also in agreement with distance estimates of HVCs
      and IVCs (Wakker \cite{Wak04}).
    
      These differences in the apparent importance of the cold cores in
      HVCs and IVCs, may help to distinguish between HVCs and IVCs. For
      example, Fig.~\ref{Fig06} illustrates the sky distribution of
      Gaussians with intermediate velocities. In Fig.~\ref{Fig06}a, we
      observe a concentration of points at $l \approx 300\degr$,
      $-90\degr < b < -65\degr$. In Fig.~\ref{Fig06}b, we only see small
      concentrations of weak Gaussians in this region. This is typical
      of HVCs, and we know that the discussed concentration of
      relatively low velocity is a part of the Magellanic Stream. In
      general, the velocities in the Magellanic Stream are high, but
      Fig.~\ref{Fig06} presents only a small part of the full stream. In
      this part the velocity is relatively low, due to projection
      effects.
    
      A more interesting example is found in Figs.~\ref{Fig03}a and
      \ref{Fig05}a. In Fig.~\ref{Fig03}a (line-width group 2), we find
      two weak concentrations of dots at $(l, b) = (120\degr, 29\degr)$
      and $(110\degr, 34\degr)$. In Fig.~\ref{Fig05}a (line-width group
      1), these clouds are considerably brighter, which is a
      characteristic of IVCs. The mean LSR velocities of the objects are
      $-80.9$ and $-78.2~\mathrm{km\,s}^{-1}$, respectively. These
      velocities are slightly above the limit $|V_\mathrm{C}| =
      74~\mathrm{km\,s}^{-1}$, which we use to separate IVCs from HVCs. 
      There is usually, however, no sharp separation between statistical
      distributions. The velocity distributions of HVCs and IVCs
      probably partially overlap, as do the line-width distributions of
      Gaussians from the first and second line-width groups. Based on
      their core-envelope structure, we expect that the two clouds are,
      in fact, intermediate-velocity clouds with velocities that are
      characteristic of HVCs. The same is possibly true for the cloud
      around $(l, b, V_C) = (39\degr, 63.5\degr,
      -80.4~\mathrm{km\,s}^{-1})$, in Figs.~\ref{Fig03}a and
      \ref{Fig05}a. The Magellanic Stream, however, is classified as an
      HVC, even though, in one part its velocities are more
      characteristic of IVCs.

      In this context, it is interesting to consider the Outer Arm
      Cloud. From the comparison of our sky distribution images, it
      appears in most parts that the properties of the gas in the OAC
      correspond to those of the HVC. In two regions only, this large
      HVC complex could be overlaid with a considerable amount of IVC
      gas. One of these regions is located at about $l = 100\degr$, and
      the corresponding IVC gas appears to form an extension of the high
      latitude IV Arch, to close to zero latitude. Another region with a
      large number of intermediate-velocity cores, located at $(l, b) =
      (179\degr, 15\degr)$, may for example be the northern extension of
      the AC Shell. In any case, the gas properties of the OAC appear to
      be different from those of normal gas in the Galactic disc.
    
      Kalberla \& Haud (\cite{Kal06}) found that the cold cores of HVC
      complexes have a random velocity distribution, with a typical
      dispersion of $20~\mathrm{km\,s}^{-1}$. Kuntz \& Danly
      (\cite{Kun96}) found evidence that the infall velocity of IVCs
      reaches a maximum at positions of highest column density. Similar
      results can be seen in our Fig.~\ref{Fig01}: at negative
      velocities the line-width group-1 Gaussians, in the intermediate
      velocity range, reach a maximum in their frequency distribution at
      $(V_\mathrm{C}, \mathrm{FWHM}) = (-63.5~\mathrm{km\,s}^{-1},
      7.9~\mathrm{km\,s}^{-1})$, whereas the components of line-width
      group 2 have a corresponding maximum at $(V_\mathrm{C},
      \mathrm{FWHM}) = (-49.4~\mathrm{km\,s}^{-1},
      22.4~\mathrm{km\,s}^{-1})$, i.e., the cold cores often have larger
      infall velocities than their warmer envelopes.

      In Fig.~\ref{Fig01}, the distribution of group-1 Gaussians is
      contaminated by spurious components from profiles close to the
      Galactic plane. To consider this further, we made similar figures
      for the regions $|b| > b_\mathrm{lim}$, with different
      $b_\mathrm{lim} \ge 15\degr$. In these figures, the frequency
      maximum of group-1 Gaussians, in the intermediate velocity range,
      is shifted to lower velocities, in most cases close to
      $V_\mathrm{C} = -54~\mathrm{km\,s}^{-1}$. The location of the
      frequency maxima of group-2 Gaussians remains practically
      unchanged. This indicates that the velocity differences are only
      partly explained by the contamination.

      At positive velocities, it is difficult to draw a similar
      conclusion. From symmetry considerations however, it is
      interesting to mention that there is a weak density enhancement at
      $(V_\mathrm{C}, \mathrm{FWHM}) = (74.9~\mathrm{km\,s}^{-1},
      7.5~\mathrm{km\,s}^{-1})$, but for higher values of
      $b_\mathrm{lim}$, it disappears completely from the plots. At the
      same time, the density enhancement, corresponding to the envelopes
      of the positive velocity IVCs, becomes much more prominent in
      higher latitude plots, than in Fig.~\ref{Fig01}.
    
   \section{Conclusions}

      Different and almost arbitrary definitions of the IVCs and HVCs
      exist so far, and the main HVC catalogues (Hulsbosch \& Wakker
      \cite{Hul88}, Morras et al. \cite{Mor00}) have been compiled by
      visual inspection of the observed profiles for the presence of
      components at high velocities. In this paper, we have tried an
      automated approach by decomposing all profiles into Gaussian
      components and studying the frequency distributions of the
      parameters of the obtained Gaussians. Our approach identifies most
      well-known HVC and IVC complexes on the basis of remarkable
      density enhancements in particular regions of the ($V_\mathrm{C} -
      \mathrm{FWHM}$) frequency diagram of the Gaussian parameters. The
      general properties of objects, separated in this way are almost
      identical to the properties of IVCs and HVCs, defined in the
      traditional way:
      \begin{itemize}
         \item the sky distributions of $T_\mathrm{b}$ and
            $V_\mathrm{LSR}$ of statistically-identified IVCs and HVCs,
            are similar to those of the classical complexes;
         \item the division line between HVCs and IVCs most likely lies
            at about $|V_\mathrm{LSR}| = 74~\mathrm{km\,s}^{-1}$;
         \item both HVCs and IVCs have a two-component core-envelope
            structure;
         \begin{itemize}
            \item the cold cores of HVCs and IVCs have almost the same
               mean line-widths: $\mathrm{FWHM} \approx 7.3$ and
               $7.2~\mathrm{km\,s}^{-1}$, respectively;
            \item when the mean line-width of the envelopes of HVCs
               is $\mathrm{FWHM} \approx 27.2~\mathrm{km\,s}^{-1}$, the
               corresponding value for IVCs is only
               $22.2~\mathrm{km\,s}^{-1}$;
            \item the angular dimensions and the brightnesses of HVC
               cores, derived using the LAB, are smaller than the
               corresponding values for the cores of IVCs.
            \item the infall velocities of the cores of IVCs often
               appear to be higher than the infall velocities of the
               corresponding envelopes.
         \end{itemize}
      \end{itemize}
      The differences in the core-envelope structure of the HVCs and
      IVCs can provide additional information to help distinguish
      between the two cloud types.

      Considering the frequency distribution, presented in
      Fig.~\ref{Fig01}, and other properties of IVCs and HVCs discussed
      in this paper, the suggestion by Kuntz \& Danly (\cite{Kun96}, see
      also Kerp et al. \cite{Ker96}, \cite{Ker99}) that IVCs were once
      HVCs that have decelerated on approach to the disc, appears
      plausible. In Fig.~\ref{Fig01}, a deep minimum is present between
      maxima, which correspond to negative-velocity IVCs and HVCs. This
      minimum could indicate that most observable intermediate-velocity
      clouds are the results of a large accretion event, or that the
      lifetime of IVCs and HVCs, is much longer than the interaction
      phase (deceleration time) between high-velocity clouds and the
      Galactic disc, or halo gas.

      At the same time, it is commonly believed that the metallicities
      of HVCs are lower than those of IVCs (Wakker \cite{Wak01}), which
      makes the possibility of a relationship between high- and
      intermediate-velocity cloud complexes controversial. This
      metallicity difference could apparently be explained if IVCs were
      HVCs at later stages of the HVC -- disc collision. The clouds
      would then have accreted a large amount of Galactic matter. From
      numerical simulations, however, remains unclear the extent to
      which such gas-mixing in cloud cores can occur (e.g. 
      Santill\'{a}n et al \cite{San99}, but see also Vieser \cite{Vie01}).

      A more detailed discussion of the properties of IVCs and HVCs,
      would require not only knowledge of the probability with which a
      particular Gaussian belongs to a intermediate- or high-velocity
      gas cloud, but a full identification of structures in position --
      position -- velocity -- line width space.
    
   \begin{acknowledgements}
      The author would like to thank W.~B.~Burton for providing the
      preliminary data from the LDS for program testing prior the
      publication of the survey. A considerable part of the work on
      creating the decomposition program was done during the stay of
      U.~Haud at the Radioastronomical Institute of Bonn University (now
      Argelander-Institut f\"ur Astronomie). The hospitality of the
      staff members of the Institute is greatly appreciated. We thank
      the anonymous referee and the A\&A editor M. Walmsley for fruitful
      discussions and considerable help. The project was supported by
      the Estonian Science Foundation grant no. 6106.
   \end{acknowledgements}

\end{document}